\documentclass[12pt]{article}
\usepackage{amssymb,amsmath,cite,epsfig}
\usepackage[top=2.5cm, bottom=2.5cm, left=2.5cm, right=2.5cm]{geometry}

\input{tcilatex}
\begin{document}

\begin{titlepage}
\vspace*{3cm}
\begin{center}
{\Large \textsf{\textbf{ Noncommutative of  space-time and the relativistic  Hydrogen atom }}}
\end{center}
\vskip 5mm
\begin{center}
{\large \textsf{Slimane Zaim,  and Yazid Delenda}}\\
\vskip 5mm
D\'{e}partement de Physique, Facult\'{e} des Sciences,\\
Universit\'{e} Hadj Lakhdar -- Batna, Algeria.\\
\end{center}
\vskip 2mm
\begin{center}{\large\textsf{\textbf{Abstract}}}\end{center}
\begin{quote}
We  study  the Klein-Gordon equation in a
non-commutative space-time as applied to the Hydrogen atom to
extract the energy levels, by considering the second-order
corrections in the non-commutativity parameter and by comparing to the 2S - 1S transition
 accuracy we get a bound on the parameter of noncommutativity. Phenomenologically
we show that non-commutativity is the source of lamb shift
corrections and spin electron.
\end{quote}
\vspace*{2cm}


\noindent\textbf{\sc Pacs numbers}: 11.10.Nx, 81Q05, 03.65.-w
\end{titlepage}

\section{Introduction}

The noncommutativity of spatial rotations in three and more dimensions is an
idea which is deeply ingrained in our theories. The non-commutativity has
received a wide appreciation as an alternative approach to understanding
many physical phenomenon such as the ultraviolet and infrared divergences 
\cite{1}, unitarity violation \cite{2}, causality \cite{3}, and new physics
at very short distances of the Planck-length order \cite{4,41,42}.

The non-commutative field theory is motivated by the natural extension of
the usual quantum mechanical commutation relations between position and
momentum, by imposing further commutation relations between position
coordinates themselves. As in usual quantum mechanics, the non-commutativity
of position coordinates immediately implies a set of uncertainty relations
between position coordinates analogous to the Heisenberg uncertainty
relations between position and momentum; namely: 
\begin{equation}
\left[ \hat{x}^{\mu },\hat{x}^{\nu }\right] _{\ast }=i\theta ^{\mu \nu },
\end{equation}%
where $\hat{x}^{\mu }$ are the coordinate operators and $\theta ^{\mu \nu }$
are the non-commutativity parameters of dimension of area that signify the
smallest area in space that can be probed in principle.We use the symbol $%
\ast $ in equation (1) to denote the product of the non-commutative
structure. This idea is similar to the physical meaning of the Plank
constant in the relation $\left[ \hat{x}_{i},\hat{p}_{j}\right] =i\hbar
\delta _{ij}$, which as is known is the smallest phase-space in quantum
mechanics. The space-time noncommutativity has gained considerable interest
in current literature and a search for any possible upper bound on the
time-space NC parameter using recent experimental feedback is very much
desirable. The issue of space-time noncommutativity is worth pursuing in its
own right because of its deep connection with such fundamental notions as
unitarity and causality. It was argued that introduction of space-time
noncommutativity spoils unitarity or even causality. Much attention has been
devoted in recent times to circumvent these difficulties in formulating
theories with $\theta ^{0\nu }\neq 0$ \cite{32,33,34,35}. We do not consider
momentum space NC effects as have been done by \cite{36}. Our motivation is
to study the effect of noncommutativity on the level of quantum mechanics
when time-space noncommutativity is accounted for. One can study the
physical consequences of this theory by making detailed analytical estimates
for measurable physical quantities and compare the results with experimental
data to find an upper bound on the $\theta $ parameter. The most obvious
natural phenomena to use in hunting for non-commutative effects are simple
quantum mechanics systems, such as the hydrogen atom \cite{5,6,7}. In the
non-commutative space-time one expects the degeneracy of the initial
spectral line to be lifted, thus one may say that non-commutativity plays
the role of spin.

In this work we present an important contribution to the non-commutative
approach to the hydrogen atom. Our goal is to solve the Klein Gorden
equation for the Coulomb potential in a non-commutative space-time up to
second-order of the non-commutativity parameter using the Seiberg-Witten
maps and the Moyal product. We thus find the non-commutative modification of
the energy levels of the hydrogen atom and we show that the
non-commutativity is the source of spin as a result of rotation of
space-time dimensions.

This paper is organized as follows. In section 2 we derive the corresponding
Seiberg-Witten maps up to the second order of $\theta $ for the various
dynamical fields, and we propose an invariant action of the non-commutative
charged scalar field in the presence of an electric field. In section 3,
using the generalised Euler-Lagrange field equation, we derive the deformed
Klein-Gordon (KG) equation. Applying these results to the hydrogen atom, we
solve the deformed KG equation and obtain the non-commutative modification
of the energy levels. The last section is devoted to a discussion.

\section{Seiberg-Witten maps}

Here we look for a mapping $\phi ^{A}\rightarrow \hat{\phi}^{A}$ and $%
\lambda \rightarrow \hat{\lambda}\left( \lambda ,A_{\mu }\right) $, where $%
\phi ^{A}=(A_{\mu },\varphi )$ is a generic field, $A_{\mu }$ and $\varphi $
are the gauge and charged scalar fields respectively (the Greek and Latin
indices denote curved and tangent space-time respectively), and $\lambda $
is the $\mathrm{U}(1)$ gauge Lie-valued infinitesimal transformation
parameter, such that: 
\begin{equation}
\hat{\phi}^{A}\left( A\right) +\hat{\delta}_ {\hat{\lambda}} \hat{\phi}
^{A}\left( A\right) =\hat{\phi}^{A}\left( A+\delta _{\lambda }A\right),
\label{eq:trans}
\end{equation}
where $\delta _{\lambda }$ is the ordinary gauge transformation and $\hat{
\delta}_{\hat{\lambda}}$ is a noncommutative gauge transformation which are
defined by: 
\begin{eqnarray}  \label{eq:tempo1}
\hat{\delta}_{\hat{\lambda}}\hat{\varphi} =i\hat{\lambda}\ast \hat{\varphi}%
,&\qquad&\delta _{\lambda }\varphi =i\lambda \varphi, \\
\hat{\delta}_{\hat{\lambda}}\hat{A}_{\mu } =\partial _{\mu }\hat{\lambda}+i %
\left[ \hat{\lambda},\hat{A}_{\mu }\right] _{\ast },&\qquad&\delta _{\lambda
}A_{\mu }=\partial _{\mu}\lambda.  \label{tempo2}
\end{eqnarray}

In accordance with the general method of gauge theories, in the
non-commutative space-time, using these transformations one can get at
second order in the non-commutative parameter $\theta ^{\mu \nu }$ (or
equivalently $\theta $) the following Seiberg--Witten maps \cite{8}: 
\begin{eqnarray}
\hat{\varphi} &=&\varphi +\theta \varphi ^{1}+\theta ^{2}\varphi ^{2}+%
\mathcal{O}\left( \theta ^{3}\right) , \\
\hat{\lambda} &=&\lambda +\theta \lambda ^{1}\left( \lambda ,A_{\mu }\right)
+\theta ^{2}\lambda ^{2}\left( \lambda ,A_{\mu }\right) +\mathcal{O}\left(
\theta ^{3}\right) , \\
\hat{A}_{\xi } &=&A_{\xi }+\theta A_{\xi }^{1}\left( A_{\xi }\right) +\theta
^{2}A_{\xi }^{2}\left( A_{\xi }\right) +\mathcal{O}\left( \theta ^{3}\right)
,  \label{eq:SWM1} \\
\hat{F}_{\mu \xi } &=&F_{\mu \xi }\left( A_{\xi }\right) +\theta F_{\mu \xi
}^{1}\left( A_{\xi }\right) +\theta ^{2}F_{\mu \xi }^{2}\left( A_{\xi
}\right) +\mathcal{O}\left( \theta ^{3}\right) ,  \label{eq:SWM2}
\end{eqnarray}%
where 
\begin{equation}
F_{\mu \nu }=\partial _{\mu }A_{\nu }-\partial _{\nu }A_{\mu }.
\end{equation}

To begin, we consider a non-commutative field theory with a charged scalar
particle in the presence of an electrodynamic gauge field in a Minkowski
space-time. We can write the action as: 
\begin{equation}
\mathcal{S}=\int d^{4}x\,\left(\eta^{\mu\nu}\left(\hat{D}_\mu\hat{ \varphi}
\right)^\dagger\ast\hat{D}_\nu\hat{\varphi}+m^{2}\hat{ \varphi}^\dagger\ast 
\hat{\varphi}-\frac{1}{4}\hat{F}_{\mu\nu}\ast \hat{F}^{\mu \nu}\right),
\label{eq:action}
\end{equation}
where the gauge covariant derivative is defined as: $\hat{D}_\mu \hat{%
\varphi }=\left(\partial_\mu+ie\hat{A}_\mu\right)\ast \hat{\varphi}$.

Next we use the generic-field infinitesimal transformations \eqref{eq:tempo1}
and \eqref{tempo2} and the star-product tensor relations to prove that the
action in eq. \eqref{eq:action} is invariant. By varying the scalar density
under the gauge transformation and from the generalised field equation and
the Noether theorem we obtain \cite{9}: 
\begin{equation}
\frac{\partial \mathcal{L}}{\partial \hat{\varphi}}-\partial _{\mu }\frac{%
\partial \mathcal{L}}{\partial \left( \partial _{\mu }\hat{\varphi}\right) }%
+\partial _{\mu }\partial _{\nu }\frac{\partial \mathcal{L}}{\partial \left(
\partial _{\mu }\partial _{\nu }\hat{\varphi}\right) }-\partial _{\mu
}\partial _{\nu }\partial _{\sigma }\frac{\partial \mathcal{L}}{\partial
\left( \partial _{\mu }\partial _{\nu }\partial _{\sigma }\hat{\varphi}%
\right) }+\mathcal{O}\left( \theta ^{3}\right) =0.  \label{eq:field}
\end{equation}

\section{ Non-commutative Klein-Gordon equation}

In this section we study the Klein-Gordon equation for a Coulomb interaction
($-e/r$) in the free non-commutative space-time. This means that we will
deal with solutions of the U$(1)$ gauge-free non-commutative field equations 
\cite{10}. For this we use the modified field equations in eq. %
\eqref{eq:field} and the generic field $\hat{A}_{\mu }$ so that: 
\begin{equation}
\delta \hat{A}_{\mu }=\partial _{\mu }\hat{\lambda}-ie\hat{A}_{\mu }\ast 
\hat{\lambda}+ie\hat{\lambda}\ast \hat{A}_{\mu },
\end{equation}%
and the free non-commutative field equations: 
\begin{equation}
\partial ^{\mu }\hat{F}_{\mu \nu }-ie\left[ \hat{A}^{\mu },\hat{F}_{\mu \nu }%
\right] _{\ast }=0.  \label{eq:freefield}
\end{equation}%
Using the Seiberg-Witten maps \eqref{eq:SWM1}--\eqref{eq:SWM2} and the
choice \eqref{eq:freefield}, we can obtain the following deformed Coulomb
potential \cite{10}: 
\begin{equation}
\hat{a}_{0}=-\frac{e}{r}-\frac{e^{3}}{\,r^{4}}\theta ^{0j}x_{j}+\frac{e^{5}}{%
2\,r^{5}}\left[ \left( \theta ^{0j}\right) ^{2}-5\left( \frac{\theta
^{0j}x_{j}}{r}\right) ^{2}\right] +\mathcal{O}\left( \theta ^{3}\right) ,
\end{equation}%
Using the modified field equations in eq. \eqref{eq:field} and the generic
field $\hat{\varphi}$ so that: 
\begin{equation}
\delta _{\hat{\lambda}}\hat{\varphi}=i\hat{\lambda}\ast \hat{\varphi},
\end{equation}%
the Klein-Gordon equation in a non-commutative space-time in the presence of
the vector potential $\hat{A}_{\mu }$ can be cast into: 
\begin{equation}
\left( \eta ^{\mu \nu }\partial _{\mu }\partial _{\nu }-m^{2}\right) \hat{%
\varphi}\,+\left( ie\eta ^{\mu \nu }\partial _{\mu }\hat{A}_{\nu }-e^{2}\eta
^{\mu \nu }\hat{A}_{\mu }\ast \hat{A}_{\nu }+2ie\eta ^{\mu \nu }\hat{A}_{\mu
}\partial _{\nu }\right) \hat{\varphi}=0.  \label{eq:KGmod}
\end{equation}

Now using the fact that: 
\begin{equation}
\eta ^{\mu \nu }\partial _{\mu }\partial _{\nu }=-\partial _{0}^{2}+\Delta ,
\end{equation}%
and 
\begin{equation}
2ie\eta ^{\mu \nu }\hat{A}_{\mu }\partial _{\nu }=i\frac{2e^{2}}{r}\partial
_{0}+2i\frac{e^{4}}{\,r^{4}}\theta ^{0j}x_{j}\partial _{0}-i\frac{e^{6}}{%
\,r^{5}}\left[ \left( \theta ^{0j}\right) ^{2}-5\left( \frac{\theta
^{0j}x_{j}}{r}\right) ^{2}\right] \partial _{0},
\end{equation}%
and 
\begin{equation}
-e^{2}\eta ^{\mu \nu }\hat{A}_{\mu }\ast \hat{A}_{\nu }=\frac{e^{4}}{r^{2}}+2%
\frac{e^{6}}{\,r^{5}}\theta ^{0j}x_{j}+\frac{e^{8}}{\,r^{6}}\left[ \left(
\theta ^{0j}\right) ^{2}-4\left( \frac{\theta ^{0j}x_{j}}{r}\right) ^{2}%
\right] ,
\end{equation}%
then the Klein-Gordon equation \eqref{eq:KGmod} up to $\mathcal{O}\left(
\theta ^{3}\right) $ takes the form:

\begin{multline}
\left[ -\partial _{0}^{2}+\Delta -m_{e}^{2}+\frac{e^{4}}{r^{2}}+i\frac{2e^{2}%
}{r}\partial _{0}+2i\frac{e^{4}}{\,r^{4}}\theta ^{0j}x_{j}\partial _{0}-i%
\frac{e^{6}}{\,r^{5}}\left[ \left( \theta ^{0j}\right) ^{2}-5\left( \frac{%
\theta ^{0j}x_{j}}{r}\right) ^{2}\right] \partial _{0}\right.  \notag \\
\left. +2\frac{e^{6}}{\,r^{5}}\theta ^{0j}x_{j}+\frac{e^{8}}{\,r^{6}}\left[
\left( \theta ^{0j}\right) ^{2}-4\left( \frac{\theta ^{0j}x_{j}}{r}\right)
^{2}\right] \right] \hat{\varphi}=0.  \label{eq:temp1}
\end{multline}
The solution to eq. \eqref{eq:temp1} in spherical polar coordinates $%
(r,\theta ,\phi )$ takes the separable form \cite{11}: 
\begin{equation}
\hat{\varphi}(r,\theta ,\phi ,t)=\frac{1}{r}\hat{R}(r)\hat{Y}\left( \theta
,\phi \right) \exp (-iEt).
\end{equation}%
Then eq. \eqref{eq:temp1} reduces to the radial equation: 
\begin{multline}
\left[ \frac{d^{2}}{dr^{2}}-\frac{l(l+1)-e^{4}}{r^{2}}+\frac{2Ee^{2}}{r}%
+E^{2}-m_{e}^{2}+2E\frac{e^{4}}{\,r^{4}}\theta ^{0j}x_{j}+2\frac{e^{6}}{%
\,r^{5}}\theta ^{0j}x_{j}\right.  \notag \\
\left. -E\frac{e^{6}}{\,r^{5}}\left[ \left( \theta ^{0j}\right) ^{2}-5\left( 
\frac{\theta ^{0j}x_{j}}{r}\right) ^{2}\right] +\frac{e^{8}}{\,r^{6}}\left[
\left( \theta ^{0j}\right) ^{2}-4\left( \frac{\theta ^{0j}x_{j}}{r}\right)
^{2}\right] \right] \hat{R}(r)=0.  \label{eq:radial}
\end{multline}

In eq. \eqref{eq:radial} the coulomb potential in non-commutative space-time
appears within the perturbation terms: 
\begin{equation}
H_{\text{pert}}^{\theta }=2E\frac{e^{4}}{\,r^{4}}\theta ^{0j}x_{j}+2\frac{%
e^{6}}{\,r^{5}}\theta ^{0j}x_{j}+5E\frac{e^{6}}{\,r^{5}}\left( \frac{\theta
^{0j}x_{j}}{r}\right) ^{2}-E\frac{e^{6}}{\,r^{5}}\left( \theta ^{0j}\right)
^{2}-4\frac{e^{8}}{\,r^{6}}\left( \frac{\theta ^{0j}x_{j}}{r}\right) ^{2}+%
\frac{e^{8}}{\,r^{6}}\left( \theta ^{0j}\right) ^{2},
\label{eq:perturbation}
\end{equation}%
where the first term is the the electric dip\^{o}le-dip\^{o}le iteraction
created by the non-commutativity, the second is the electric dip\^{o}%
le-quadrup\^{o}le iteractions, \ the third and fourth terms are the electric
quadrup\^{o}le-quadrup\^{o}le iteraction, and the last terms are similar to
the Van-der--Waals potential energy forces between two atoms. These
interactions show us that the effect of space-time non-commutativity on the
interaction of the electron and the proton is equivalent to an extention of
two nucleus interactions at a considerable distance. This idea effectively
confirms the presence of gravity at this level. To investigate the
modification of the energy levels by (26) , we use the first-order
perturbation theory. The spectrum of $H_{0}$ and the corresponding wave
functions are well known and given by:

\begin{equation}
R_{nl}(r)=\sqrt{\frac{a}{n+\nu +1}}\left( \frac{n!}{\Gamma \left( n+2\nu
+2\right) }\right) ^{1/2}x^{\nu +1}e^{-x/2}L_{n}^{2\nu +1}(x)\,,
\end{equation}%
where the relativistic energy levels are given by:

\begin{equation}
E=E_{n,l}=\frac{m_{e}\left( n+\frac{1}{2}+\sqrt{\left( l+\frac{1}{2}\right)
^{2}-\alpha ^{2}}\right) }{\left[ \left( n+\frac{1}{2}\right) ^{2}+\left( l+%
\frac{1}{2}\right) ^{2}+2\left( n+\frac{1}{2}\right) \sqrt{\left( l+\frac{1}{%
2}\right) ^{2}-\alpha ^{2}}\right] ^{\frac{1}{2}}}\,\qquad .
\end{equation}%
and $L_{n}^{2\nu +1}$ are the associated Laguerre polynomials [20], with the
following notations:

\begin{equation}
\nu =-\frac{1}{2}+\sqrt{\left( l+\frac{1}{2}\right) ^{2}-\alpha ^{2}},\text{
\ \ \ \ \ \ \ \ \ }\alpha =e^{2},\text{\ \ \ \ \ \ \ \ \ \ \ \ \ \ }a=\sqrt{%
m_{e}^{2}-E^{2}}
\end{equation}

\section{Noncommutative corrections of the energy}

Now to obtain the modification to the energy levels as a result of the terms %
\eqref{eq:perturbation} due to the non-commutativity of space-time we use
perturbation theory. For simplicity, first of all, we take $\theta
_{i}=\delta _{i3}\theta $ $\left( \theta ^{0j}x_{j}=\theta r\cos \vartheta
\right) $ and assume that the other components are all zero and also the
fact that in the first-order perturbation theory the expectation value of $%
\cos \vartheta /r^{3}$,$\cos \vartheta /r^{4},\cos \vartheta ^{2}/r^{5},$ $%
\cos \vartheta ^{2}/r^{6}$, $1/r^{5}$ and $1/r^{6}$ are asfollows :

\begin{equation}
\langle \psi _{nlm}^{0}+\psi ^{1}\left\vert H_{\text{pert}}^{\theta \left(
1\right) }+H_{\text{pert}}^{\theta \left( 2\right) }\right\vert \psi
^{0}+\psi ^{1}\rangle
\end{equation}%
where

\begin{equation*}
H_{\text{pert}}^{\theta \left( 1\right) }=\theta 2e^{4}\left( E\frac{\cos
\vartheta }{\,r^{3}}+e^{2}\frac{\cos \vartheta }{\,r^{4}}\right)
\end{equation*}

\begin{equation*}
H_{\text{pert}}^{\theta \left( 2\right) }=\theta ^{2}e^{6}\left( 5E\frac{%
\cos \vartheta ^{2}}{\,r^{5}}-4e^{2}\frac{\cos \vartheta ^{2}}{\,r^{6}}-E%
\frac{1}{\,r^{5}}+e^{2}\frac{1}{\,r^{6}}\right) ,
\end{equation*}%
and

\begin{equation*}
\psi _{nlm}^{0}=R_{nl}(r)Y_{l}^{m}(\theta ,\phi )
\end{equation*}

\begin{equation*}
\psi ^{1}=\dsum \frac{\langle \psi _{nlm}^{0}\left\vert H_{\text{pert}%
}^{\theta \left( 1\right) }\right\vert \psi _{nkm}^{0}\rangle }{E_{nl}-E_{nk}%
}\psi _{nkm}^{0}
\end{equation*}

By taking into account the fact that

\begin{eqnarray*}
\langle \psi _{nlm}^{0} &\mid &\cos \vartheta \mid \psi _{nl^{\prime
}m^{\prime }}^{0}\rangle =B_{l^{\prime }+1}^{m^{\prime }}\delta
_{l,l^{\prime }+1}\delta _{m,m^{\prime }}+B_{l^{\prime }}^{m^{\prime
}}\delta _{l,l^{\prime }-1}\delta _{m,m^{\prime }} \\
\langle \psi _{nlm}^{0} &\mid &\cos \vartheta ^{2}\mid \psi _{nl^{\prime
}m^{\prime }}^{0}\rangle =B_{l^{\prime }+1}^{m^{\prime }}B_{l^{\prime
}+2}^{m^{\prime }}\delta _{l,l^{\prime }+2}\delta _{m,m^{\prime
}}+B_{l^{\prime }}^{m^{\prime }}B_{l^{\prime }-1}^{m^{\prime }}\delta
_{l^{\prime }-2}\delta _{m,m^{\prime }}+ \\
&&\left[ B_{l^{\prime }+1}^{m^{\prime }}B_{l^{\prime }+1}^{m^{\prime
}}+B_{l^{\prime }}^{m^{\prime }}B_{l^{\prime }}^{m^{\prime }}\right] \delta
_{l^{\prime }l}\delta _{mm^{\prime }}
\end{eqnarray*}

where

\begin{equation*}
B_{l}^{m}=\sqrt{\frac{\left( l+m\right) \left( l-m\right) }{\left(
2l+1\right) \left( 2l-1\right) }}
\end{equation*}

we have

\begin{equation*}
\psi ^{1}=\theta ^{2}\frac{e^{4}B_{l}^{m}}{E_{nl}-E_{nl-1}}\left[
E_{nl}f(3)+e^{2}f(4)\right] \psi _{nl-1m}^{0}+\theta ^{2}\frac{%
e^{4}B_{l+1}^{m}}{E_{nl}-E_{nl+1}}\left[ E_{nl}f(3)+e^{2}f(4)\right] \psi
_{nl+1m}^{0}
\end{equation*}

where $f(k)=\left\langle \frac{1}{r^{k}}\right\rangle $ and completely
defined by 
\begin{eqnarray}
\langle nlm\mid r^{-k}\mid nlm\rangle &=&\int_{0}^{\infty
}R_{nl}^{2}(r)r^{-k}dr  \notag  \label{eq:expectationvalue} \\
&=&\frac{2^{k-1}a^{k}n!}{\left( n+\nu +1\right) \Gamma \left( n+2\nu
+2\right) }\int_{0}^{\infty }x^{2\nu +2-k}e^{-x}\left[ L_{n}^{2\nu +1}(x)%
\right] ^{2}dx  \notag \\
&=&f(k)\qquad k=3,4,5,6.
\end{eqnarray}

We use the relation between the confluent hypergeometric function $F(-n;\nu
+1;x)$ and the associated Laguerre polynomials $L_{n}^{\nu }(x)$, namely: 
\begin{equation}
L_{n}^{\nu }(x)=\frac{\Gamma \left( n+\nu +1\right) }{\Gamma \left(
n+1\right) \Gamma \left( \nu +1\right) }F(-n;\nu +1;x),
\end{equation}%
\begin{multline}
\int_{0}^{\infty }x^{\nu -1}e^{-x}\left[ F(-n;\gamma ;x)\right] ^{2}dx=\frac{%
n!\Gamma (\nu )}{\gamma \left( \gamma +1\right) \cdots \left( \gamma
+n-1\right) }\left\{ 1+\frac{n\left( \gamma -\nu -1\right) \left( \gamma
-\nu \right) }{1^{2}\gamma }\right. + \\
\left. +\frac{n\left( n-1\right) \left( \gamma -\nu -2\right) \left( \gamma
-\nu -1\right) \left( \gamma -\nu \right) \left( \gamma -\nu +1\right) }{%
1^{2}2^{2}\gamma \left( \gamma +1\right) }+\cdots \right. \\
\left. \cdots +\frac{n\left( n-1\right) \cdots 1\left( \gamma -\nu -n\right)
\cdots \left( \gamma -\nu +n-1\right) }{1^{2}2^{2}\cdots n^{2}\gamma \left(
\gamma +1\right) \cdots \left( \gamma +n-1\right) }\right\} .
\end{multline}%
Equation \eqref{eq:expectationvalue} becomes: 
\begin{eqnarray}
\langle nlm\mid r^{-3}\mid nlm^{\prime }\rangle &=&\int_{0}^{\infty
}R_{nl}^{2}(r)r^{-3}dr\delta _{mm^{\prime }}  \notag \\
&=&\frac{4a^{3}n!}{\left( n+\nu +1\right) \Gamma \left( n+2\nu +2\right) }%
\int_{0}^{\infty }x^{2\nu -1}e^{-x}\left[ L_{n}^{2\nu +1}(x)\right]
^{2}dx\delta _{mm^{\prime }}  \notag \\
&=&\frac{4a^{3}n!}{\left( n+\nu +1\right) \Gamma \left( n+2\nu +2\right) }%
\left[ \frac{\Gamma \left( n+2\nu +2\right) }{\Gamma \left( n+1\right)
\Gamma \left( 2\nu +2\right) }\right] ^{2}\times  \notag \\
&&\qquad \qquad \qquad \qquad \times \int_{0}^{\infty }x^{2\nu -1}e^{-x} 
\left[ F(-n;2\nu +2;x)\right] ^{2}dx\delta _{mm^{\prime }}  \notag \\
&=&\frac{2a^{3}}{\nu \left( 2\nu +1\right) \left( n+\nu +1\right) }\left\{ 1+%
\frac{n}{\left( \nu +1\right) }\right\} \delta _{mm^{\prime }}=f(3),  \notag
\end{eqnarray}%
\begin{eqnarray}
\langle nlm\mid r^{-4}\mid nlm^{\prime }\rangle &=&\frac{4a^{4}}{\left( 2\nu
-1\right) \nu \left( 2\nu +1\right) \left( n+\nu +1\right) }\left[ 1+\frac{3n%
}{\left( \nu +1\right) }\right. +  \notag \\
&&\left. +\frac{3n\left( n-1\right) }{\left( \nu +1\right) \left( 2\nu
+3\right) }\right] \delta _{mm^{\prime }}  \notag \\
&=&f(4),  \notag \\
\langle nlm\mid r^{-5}\mid nlm^{\prime }\rangle &=&\frac{4a^{5}}{\left( 2\nu
-1\right) \left( \nu -1\right) \nu \left( 2\nu +1\right) \left( n+\nu
+1\right) }\left[ 1+\frac{6n}{\left( \nu +1\right) }\right. +  \notag \\
&&\left. +\frac{15n\left( n-1\right) }{\left( \nu +1\right) \left( 2\nu
+3\right) }+\frac{5n\left( n-1\right) \left( n-2\right) }{\left( \nu
+1\right) \left( 2\nu +3\right) \left( \nu +2\right) }\right] \delta
_{mm^{\prime }}  \notag \\
&=&f(5),  \notag \\
\langle nlm\mid r^{-6}\mid nlm^{\prime }\rangle &=&\frac{8a^{6}}{\left( 2\nu
-3\right) \left( 2\nu -1\right) \left( \nu -1\right) \nu \left( 2\nu
+1\right) \left( n+\nu +1\right) }\left[ 1+\frac{10n}{\left( \nu +1\right) }%
\right. +  \notag \\
&&\left. +\frac{45n\left( n-1\right) }{\left( \nu +1\right) \left( 2\nu
+3\right) }+\frac{35n\left( n-1\right) \left( n-2\right) }{\left( \nu
+1\right) \left( 2\nu +3\right) \left( \nu +2\right) }\right. +  \notag \\
&&\left. +\frac{35n\left( n-1\right) \left( n-2\right) \left( n-3\right) }{%
2\left( \nu +1\right) \left( 2\nu +3\right) \left( \nu +2\right) \left( 2\nu
+5\right) }\right] \delta _{mm^{\prime }}  \notag \\
&=&f(6).
\end{eqnarray}

The first-order correction terms are :

\begin{equation*}
\langle \psi _{nlm}^{0}\left\vert H_{\text{pert}}^{\theta \left( 1\right)
}\right\vert \psi ^{0}\rangle =0.
\end{equation*}%
Therefore, the noncommutativity of space-time to the first order has no
effect , so we study the noncommutativity effects at second order.

To second order in $\theta $, equation (14) can be written as :

\begin{equation*}
2\langle \psi ^{1}\left\vert H_{\text{pert}}^{\theta \left( 1\right)
}\right\vert \psi ^{0}\rangle +\langle \psi _{nlm}^{0}\left\vert H_{\text{%
pert}}^{\theta \left( 2\right) }\right\vert \psi ^{0}\rangle .
\end{equation*}%
So we have

\begin{equation*}
2\langle \psi ^{1}\left\vert H_{\text{pert}}^{\theta \left( 1\right)
}\right\vert \psi ^{0}\rangle =\theta ^{2}8e^{8}\left[ \frac{\left(
B_{l}^{m}\right) ^{2}}{E_{nl}-E_{nl-1}}+\frac{\left( B_{l+1}^{m}\right) ^{2}%
}{E_{nl}-E_{nl+1}}\right] \left[ E_{nl}f(3)+e^{2}f(4)\right] ^{2}
\end{equation*}%
and

\begin{equation*}
\langle \psi _{nlm}^{0}\left\vert H_{\text{pert}}^{\theta \left( 2\right)
}\right\vert \psi ^{0}\rangle =\theta ^{2}e^{6}\left\{ \left[ \left(
B_{l}^{m}\right) ^{2}+\left( B_{l+1}^{m}\right) ^{2}\right] \left[
5E_{nl}f(5)-4e^{2}f(6)\right] -E_{nl}f(5)+e^{2}f(6)\right\} .
\end{equation*}%
Putting these results together one gets

\begin{eqnarray}
\Delta E^{\mathrm{nc}}\left( n,l,m\right) &=&\theta ^{2}\alpha ^{3}\left[
8\alpha \lbrack \frac{\left( B_{l}^{m}\right) ^{2}}{E_{nl}-E_{nl-1}}+\frac{%
\left( B_{l+1}^{m}\right) ^{2}}{E_{nl}-E_{nl+1}}]\left[ E_{nl}f(3)+e^{2}f(4)%
\right] ^{2}\right.  \notag \\
&&\left. +([\left( B_{l}^{m}\right) ^{2}+\left( B_{l+1}^{m}\right) ^{2}] 
\left[ 5E_{nl}f(5)-4e^{2}f(6)\right] -E_{nl}f(5)+e^{2}f(6))\right] .
\end{eqnarray}%
The energy shift is dependent of magnetic quantum number, which clearly
reflects the existence of spin. Furthermore it is worth noting that the
correction terms containing $\theta ^{2}$ are very similar to the spin--spin
coupling, thus the non-commutative parameter $\theta $ plays the role of
spin and thus the degeneracy of levels is completly removed. The energy
levels of the hydrogen atom in the framework of the non-commutative Klein-
Gordon equation are: 
\begin{equation}
\hat{E}=E_{n,l}+\Delta E^{\mathrm{nc}}\left( n,l,m\right) .
\end{equation}

We showed that the energy-levels shift for $1S$ and $2S$ states are: 
\begin{eqnarray}
\Delta E_{1\mathrm{S}}^{\mathrm{nc}} &=&\theta ^{2}\alpha ^{3}\left(
-E_{10}f_{1\mathrm{S}}(5)+e^{2}f_{1\mathrm{S}}(6)\right) ,\text{ \ \ \ \ \ \
\ } \\
\Delta E_{2\mathrm{S}}^{\mathrm{nc}} &=&\theta ^{2}\alpha ^{3}\left(
-E_{20}f_{2\mathrm{S}}(5)+e^{2}f_{2\mathrm{S}}(6)\right) ,\text{ }
\end{eqnarray}%
One can obtain a limit for $\theta $ by comparing the corrections to
transition energies obtained using (17) with the experimental results from
hydrogen spectroscopy. We take as test levels, $1S$ and $2S$ because we have
the best experimental precision for the transition between them $[16]$:

\begin{equation*}
f_{1\mathrm{S}-2\mathrm{S}}=(2446061102474851\pm 34)HZ
\end{equation*}%
The non-commutative correction to this transition reads:

\begin{equation*}
\delta E\left( 1\mathrm{S}-2\mathrm{S}\right) =636,737\theta ^{2}\left(
Mev\right) ^{3}
\end{equation*}%
Comparing with the precision of the experimental value in (19), the bound is
given by:

\begin{equation*}
\theta \leq \left( 8Gev\right) ^{-2}
\end{equation*}%
This is in agreement with other results presented in Refs \cite{17}. Thus
the experimental signature for space-time noncommutativity differs from that
for space-space noncommutativity \cite{36}.

\section{Conclusions}

In this work we started from quantum relativistic charged scalar particle in
a canonical non-commutative space-time to find the action which is invariant
under the infinitesimal gauge transformation. By using the Seiberg-Witten
maps and the Moyal product up to second order in the non-commutativity
parameter $\theta $, we derived the deformed Klein-Gorden equation for
non-commutative Coulomb potential. By solving the deformed KG equation we
found that the energy shift up to the second order of $\theta $, is
proportional to $\theta ^{2}$, thus we explicitly accounted for spin
effects, resulting from the rotation of the time dimension in space. Hence
we can say that the Klein-Gordon equation in non-commutative space-time at
the second order of $\theta $ describes particles with spin. Thus we came to
the conclusion that the non-commutative relativistic theory degeneracy is
completely removed. This is proofs that the noncommutative space-time
responsible for a spin effects.

\end{document}